\documentclass[11pt]{amsart}

\usepackage{amsmath}
\usepackage{amssymb}
\usepackage{graphicx}
\theoremstyle{definition}

\numberwithin{equation}{section}
\title[abbreviated title ]{The Quantum Mechanics of Experiments}

\author[JF]{J\"urg Fr\"ohlich}

\address[J. Fr\"ohlich]{Department of Physics, ETH Zurich, 8093 Zurich, Switzerland}
\email{{\tt juerg@phys.ethz.ch}}

\author[AP]{Alessandro Pizzo}
\address[A. Pizzo]{Universit\`a di Roma II - Tor Vergata, Dipartimento di Matematica, 00133 Rome, Italy}
\email{\tt pizzo@mat.uniroma2.it}



\begin{document}

\begin{abstract}
This note starts with a recapitulation of what people call the ``Measurement Problem'' of Quantum Mechanics (QM).
The \textit{dissipative nature} of the quantum-mechanical time-evolution of averages of states over large ensembles 
of identical isolated systems consisting of matter interacting with the radiation field is discussed and shown to correspond 
to a \textit{stochastic} time-evolution of states of \textit{individual} systems. The importance of dissipation for the successful 
completion of measurements is highlighted. To conclude, a solution of the ``Measurement Problem'' is sketched in an 
idealized model of a double-slit experiment.
\end{abstract}

\maketitle

%%%%%%%%%%%%%%%%%%%%%%%%%%%%%%%%%%%%%%%%%%%%%%%%%%%%%%%%%%%%%%%%%%%%%%%%%%%%%%%%
\begin{center}
{\textit{Dedicated to our friends Israel Michael Sigal and Barry Martin Simon}}
\end{center}

\section{What is the ``Measurement Problem'' of QM?}\label{Intro}
In this note we summarize some work done in discussions and collaboration with Carlo Albert;
a more detailed account of our findings will appear in a paper in preparation. Among numerous problems we have debated  
is the so-called ``Measurement Problem'' of QM, in particular the question of how to come up with a logically coherent 
description of quantum-mechanical measurement processes. We have been studying such problems within a certain completion 
of QM dubbed $ETH$- Approach to (or ``$ETH$-Completion'' of) QM \cite{FS, BFS, Fr, FGP}. To give away the punchline of 
our efforts, we are convinced that, in a non-relativistic regime (with the velocity of light taken to $\infty$), the  
``Measurement Problem'' of QM can be solved in an entirely satisfactory way.

To begin with, we indicate what, in our view, the ``Measurement Problem'' of QM is, and why people have great difficulties
in solving it.\footnote{For some time-honored ideas about measurements in quantum mechanics, see also \cite{Zeh, Primas, Hepp} and references
given there.} We will, 
however, not engage in any discussion of so-called \textit{``interpretations''} of QM; our attitude towards them is summarized 
in the following verdict due to P.A.M.~Dirac: \textit{``The interpretation of quantum mechanics has been dealt with by many authors, 
and I do not want to discuss it here. I want to deal with more fundamental things.''} 

As  an example of our general approach towards developing a precise quantum theory of experiments and measurements  
an idealized model of a double-slit experiment will be discussed in Sect.~4.

\subsection{Some basic notions and premises}
In this section we summarize various well known notions, facts and premises, the purpose being to introduce convenient language
and to clarify the starting point of our analysis.

An \textit{isolated system, S,} is a physical system that has only negligibly tiny interactions with its complement (the rest of the universe).
Only for an isolated system, quantum-mechanical time evolution of operators representing physical quantities characteristic of the
system can be formulated in a universal and precise way as \textit{Heisenberg evolution} of operators acting on the Hilbert space 
of the system. In this note we will only consider isolated systems. An isolated system is \textit{open} if -- to be concrete -- it can release 
massless modes (photons and gravitons) escaping to the event horizon of the universe (or of black holes). Such modes can 
carry away energy and information that, for \textit{fundamental reasons,} cannot be retrieved in the future by any 
material devices belonging to the system. (A more general, abstract notion of open systems will be described later and is based
on what has been called the \textit{``Principle of Diminishing Potentialities,''} see \cite{Fr}.)

In this note, we focus our attention on \textit{non-relativistic} QM. We consider a system, $S$, of matter interacting with the quantized
radiation field with the property that all characteristic velocities of matter modes are tiny as compared to the velocity of light, $c$. 
In the description of such a system one may consider the limiting regime where $c\rightarrow \infty$, and this is what we will
focus on in this paper.

Let $\mathcal{H}$ denote the Hilbert space of pure state vectors of $S$. General states of $S$ are given 
by density matrices acting on $\mathcal{H}$, and physical quantities, $\widehat{X}, \widehat{Y}, \dots$, of $S$ are represented 
by selfadjoint bounded operators, $X, Y, \dots,$ on $\mathcal{H}$. Let $H_S$ denote the Hamiltonian generating the time evolution 
of operators acting on $\mathcal{H}$. The \textit{Heisenberg equation of motion} \cite{Heisenberg, Dirac} for the time-dependence 
of a selfadjoint, bounded operator, $X(t)$, representing a physical quantity, $\widehat{X}$, of $S$ at time $t\in \mathbb{R}$ is given by
\begin{equation}\label{Heisenberg}
\dot{X}(t)= \frac{i}{\hbar}\big[H_S, X(t)\big],
\end{equation}
where the dot indicates differentiation with respect to time $t$, and $\hbar$ is Planck's constant. This is a \textbf{linear deterministic} 
equation for $X(t), t\in \mathbb{R}$. It is a basic ingredient of the following analysis.

\textit{\underline{Remark}:} In equation \eqref{Heisenberg}, it is assumed that the Hamiltonian $H_S$ is a densely defined self-adjoint operator on $\mathcal{H}$.
Many important results concerning (Schr\"odinger) Hamiltonians of non-relativistic QM, e.g., proofs of
self-adjointness, etc. have been contributed by, among numerous authors, Sigal and Simon; see, e.g., \cite{Sigal, RS}.

In \textbf{Classical Mechanics}, the dynamics of an isolated system is governed by \textbf{deterministic} (most often non-linear) 
equations of motion for the \textbf{state}, $\xi(t)$, of the system as a function of time $t$. The state, $\xi$, is a point in the 
state space, $\mathfrak{X}$, of the system; ($\mathfrak{X}$ is a topological space, most often a manifold; 
in Hamiltonian mechanics, $\mathfrak{X}$ is a symplectic manifold, the \textit{phase space} of the system).
Under suitable hypotheses, the trajectory $\big\{\xi(t)\in \mathfrak{X}\,\big|\, t\in \mathbb{R}\big\}$ of the system as a function of time $t$ 
is completely determined by its initial condition $\xi(t=0)=\xi_0$. A strict \textit{Law of Causality} holds. Physical quantities (``observables'')
are described by real-valued bounded continuous functions on $\mathfrak{X}$; they have a precise value in every pure state
$\xi \in \mathfrak{X}$. Equivalently, the state space $\mathfrak{X}$ of the system can be thought of as the spectrum of the abelian algebra 
generated by all its ``observables,'' and time evolution can be viewed as given by a one-parameter family of $^{*}$-automorphisms 
of the algebra of  ``observables,'' which (under natural hypotheses) turns out to be generated by a vector field on $\mathfrak{X}$; 
in Hamiltonian mechanics, this vector field is a Hamiltonian vector field, hence generated by a real-valued (Hamilton) function 
on $\mathfrak{X}$.

In \textbf{Quantum Mechanics}, the situation is remarkably different. Physical quantities (``observables'') of a system $S$ generate a 
\textit{non-abelian} $C^{*}$-algebra. In general, they do not have precise values in pure states:\footnote{\textit{Pure states} are states that 
\textit{cannot} be represented as convex combinations of several distinct states, while \textit{mixed states} are convex combinations of 
several distinct pure states.} If $\omega$ denotes a state of $S$, 
and $X=X^{*}$ is an operator on $\mathcal{H}$ representing a physical quantity $\widehat{X}$ at an arbitrary, but fixed time, 
we define $\overline{X}^{\omega}:= \omega(X)$ to be the expectation value of $\widehat{X}$ in the state $\omega$ 
(i.e., the average of values of $\widehat{X}$ measured in a long sequence of identical experiments, with $S$ prepared 
in the state $\omega$ in every experiment), and we take
\begin{equation}\label{uncertainty}
\big(\Delta_{\omega} X\big)^{2}:= \overline{\big(X- \overline{X}^{\omega}\big)^{2}}^{\,\omega}
\end{equation}
to be the square of the \textit{uncertainty} of the value of $\widehat{X}$ in the state $\omega$. The celebrated \textit{Kochen-Specker theorem} \cite{KS}
tells us that, under some natural algebraic hypotheses, the expectation values and uncertainties of the operators representing physical quantities 
of a system $S$ in QM \textbf{cannot} be reproduced by classical random variables.\footnote{The Kochen-Specker theorem can be derived
from Gleason's theorem \cite{Gleason} and, just like the latter, requires assuming that the dimension of $\mathcal{H}$ is $\geq 3$.} Bell's
celebrated inequalities \cite{Bell} lead to the same conclusion.

Conventionally, it is claimed that, in the \textit{Heisenberg picture}, the time-dependence of an operator, $X(t)$, representing 
a physical quantitiy, $\widehat{X}$, of $S$ at time $t$ is described by the Heisenberg equations \eqref{Heisenberg}, 
while states, $\omega$, of $S$ are given by \textit{time-independent} density matrices, $\Omega$, acting on the 
Hilbert space $\mathcal{H}$ of $S$. One then goes on to claim that the Heisenberg picture is \textit{equivalent} to the 
\textit{Schr\"odinger picture}: The expectation of a physical quantity $\widehat{X}$ of $S$ at time $t$ in the state $\omega$ is given by
\begin{equation}\label{SP}
\omega \big(X(t)\big) = \text{Tr} \big[\Omega \cdot X(t)\big] = \text{Tr}\big[\Omega(t)\cdot X\big]\,, \quad X=X(0).
\end{equation}
where 
\begin{equation}\label{vN}
\dot{\Omega}(t) = -\frac{i}{\hbar} \big[H_S, \Omega(t)\big]\,, \quad \Omega(0)= \Omega\,.
\end{equation}
This is the \textit{Schr\"odinger-von Neumann equation} describing the time-depen- dence of the density matrix $\Omega(t)$ representing
the state $\omega$ of $S$ at time $t$ in the Schr\"odinger picture. We note that, under natural hypotheses (see, e.g, \cite{RS}),
equation \eqref{vN} is a \textbf{linear, deterministic} equation for the time-dependence of the density matrix $\Omega(t)$, for all times
$t\in \mathbb{R}$.

However, ever since Einstein's analysis (1917) of spontaneous and induced emission and absorption of photons by excited atoms
\cite{Einstein} and Born's analysis (1926) of collision processes \cite{Born}, it is claimed that Quantum Mechanics is 
\textbf{fundamentally probabilistic}. How can this be reconciled with the deterministic nature of the Heisenberg- and the
Schr\"odinger-von Neumann equations? This is the key question to be addressed in this note.

\subsection{The confusion introduced by the Schr\"odinger-von Neumann equation and the ``measurement postulates''
of von Neumann and L\"uders}
In text-books, the tension between the deterministic nature of the Schr\"odinger-von Neumann equation for
the quantum-mechanical time evolution of states of isolated systems and the fundamentally probabilistic nature of 
non-relativistic Quantum Mechanics usually culminates in the following two (rather confusing and
misleading) \textbf{Postulates} formulated by von Neumann and L\"uders \cite{vN, Lu}.
\begin{enumerate}
\item[I.]{The time-evolution in the Schr\"odinger picture of states of an isolated system, $S$, is described by the
time-dependent (linear, deterministic) Schr\"odinger-von Neumann equation -- \textbf{except} when some physical
quantitiy, $\widehat{X}$, characteristic of $S$ is measured, in which case the state, $\omega_i$, of $S$ immediately before
the measurement begins makes a jump to a state, $\omega_f$, occupied by the system immediately after the conclusion of
the measurement of $\widehat{X}$, with the following properties:
\begin{enumerate}
\item{the measured value, $\xi$, of $\widehat{X}$ (with $\widehat{X}$ assumed to have discrete point spectrum) coincides with the 
expectation value, $\overline{X}^{\omega_f}$, of $\widehat{X}$ in the state $\omega_f$; $\xi$ belongs to the spectrum 
of $\widehat{X}$ (= spectrum of the self-adjoint operator $X$ representing $\widehat{X}$ at the time of measurement);}
\item{the uncertainty, $\Delta_{\omega_f} X$, of the value, $\xi$, of $\widehat{X}$ in the state $\omega_f$ of $S$ \textit{vanishes}.}
\end{enumerate}
}
\item[II.]{If the uncertainty $\Delta_ {\omega_i} X$ of the value of $\widehat{X}$ in the state $\omega_i$ of $S$ immediately before the measurement
of $\widehat{X}$ begins is \textit{strictly positive} then the measured value $\xi$ of $\widehat{X}$ \textbf{cannot} be predicted; 
QM only predicts the \textbf{probability} (or frequency) of measuring the value $\xi$ in the state $\omega_i$ if the same 
measurement of $\widehat{X}$, with $S$ prepared in the same state $\omega_i$, is repeated many times. 
This probability is given by the well-known \textbf{Born Rule}.
Since it is claimed that the uncertainty $\Delta_{\omega_f} X$ of the value of $\widehat{X}$ in the state $\omega_f$ of $S$ immediately
after the conclusion of the measurement of $\widehat{X}$ \textit{vanishes}, one concludes that the two states $\omega_i$ and $\omega_f$
must be \textbf{distinct}. The stochastic transition from state $\omega_i$ to state $\omega_f$ happening in the course of a
measurement of $\widehat{X}$ is usually called \textit{``collapse of the wave function.''}
}
\end{enumerate}
These two postulates lead to the infamous \textbf{``Measurement Problem''} of QM, which is considered to be fundamental
and open by many physicists. It can be summarized in the follwoing questions and remarks.
\begin{enumerate}
\item[(i)]{What exactly characterizes a \textbf{``Measurement''}? Why is the time-evolution of the state of a system $S$ during a
``measurement'' \textbf{not} \textbf{deterministic}, more precisely \textbf{not} described by a time-dependent Schr\"odinger-von 
Neumann equation; why is it described by a stochastic jump process, i.e., a ``wave function collapse,'' satisfying Born's Rule?\footnote{See
also Bell's contribution to \cite{Bell-2}.}}
\item[(ii)]{How is it possible that the value of a physical quantity of $S$ observed in a ``measurement'' is sharp (namely 
given by an eigenvalue of the operator representing the measured quantity) if the uncertainty of the quantity in the state of $S$ 
immediately before the measurement takes place is strictly positive?}
\item[(iii)]{What causes wave functions to collapse; namely what causes the state of $S$ immediately before a measurement 
sets in to jump to a state of $S$ right after the conclusion of the measurement in which the uncertainty of the measured 
quantity vanishes?}
\item[(iv)]{What determines the \textbf{time} when a measurement begins? How long does a measurement last?}
\end{enumerate}
As far as we can tell, there aren't any good answers to these questions within something like the \textit{Copenhagen
Interpretation} of QM! One gets the impression that Postulates I and II are, at best, reasonable heuristic guidelines
for the extraction of experimentally verifiable consequences of a quantum-mechanical description of physical systems. This
does, however, not eliminate their rather confusing character! Here is what we might or should be confused about:
\begin{enumerate}
\item{``Measurements'' are physical processes, too, involving interactions between a subsystem of interest and some
(often macroscopically large) measuring devices. If the latter are taken to belong to the system considered in
our description then, to a very good approximation, this system is \textit{isolated.} Hence its time-evolution ought to be 
governed by a \textbf{deterministic} Schr\"odinger-von Neumann equation. -- Yet, apparently it isn't!}
\item{In the Copenhagen interpretation, the \textit{physical quantity} to be measured and the \textit{time} of its 
measurement are usually pre-determined by an ``observer'' (whatever this notion actually means), who is \textbf{not} part
of the system. This aspect deprives the theory of much of its predictive power, as the theory does not predict actions
of ``observers.''}
\item{If there were only \textbf{one} measurement made at ``the end of time,'' the Copenhagen ``conventional wisdom'' would be
quite satisfactory. However, we would like to describe \textit{time-ordered sequences} of many measurements made, one after
the other, on one and the same system. This leads to the well known problems caused by interference effects. Cures have been 
proposed for them; e.g., a formulation of testable predictions of QM in terms of ``consistent histories'' \cite{Griffiths, G-H}. 
But we doubt that these ideas represent a satisfactory solution of the ``measurement problem.''}
\item{In most conventional formulations (or ``interpretations'') of QM, it is assumed that measurements can be carried out in an 
arbitrarily short interval of time. By time-energy uncertainty relations, this suggests that measurements should be accompanied 
by arbitrarily large energy-fluctuations, which, in practice, they usually aren't. -- Etc.}
\end{enumerate}
If QM, and in particular quantum-mechanical processes aimed at measuring physical quantities, cannot be formulated more clearly and 
coherently than sketched here then one cannot help agreeing with P.A.M.~Dirac, who said: \textit{It seems clear 
that the present quantum mechanics is not in its final form.} 

In the remainder of this note we outline an approach to non-relativistic QM (dubbed $ETH$-Approach \cite{BFS}) that should be 
considered to be an attempt to cast QM in a \textit{final form} and, in particular, to solve the ``Measurement Problem.'' We apply
our general ideas and results to a specific example of a measurement. We expect that most of the concepts developed below 
can be extended to relativistic quantum theory \cite{Fr-2}; but many details remain to be worked out more precisely.

\section{The ``Principle of Diminishing Potentialities'' and the Dissipative Nature of the Time-Evolution of States
in Quantum Mechanics}\label{dissipation}
As announced, the following discussion is limited to non-relativistic QM. We consider a regime where the velocity
of light $c$ is taken to $\infty$; but a caricature of the quantized electromagnetic field is included implicitly in our description. 
For isolated systems, we regard the Heisenberg equations \eqref{Heisenberg} of motion of operators as fundamental, but the initial 
formulation of Matrix Mechanics \cite{Heisenberg, Dirac} (text-book QM) must be extended to a more complete framework, in order 
to enable one to describe the stochastic time-evolution of states of individual systems, including measurement processes.

In the following, it is advantageous to emphasize the concept of \textbf{``events''}, rather than talk about physical quantities
and ``observables'' (which, in the $ETH$-Approach to QM, have the status of \textit{derived quantities}). In non-relativistic QM, 
a \textbf{potential event},\footnote{more precisely, a family of ``mutually exclusive potential events''} $\mathfrak{e}$, 
is given by a partition of unity, $\mathfrak{e}:=\big\{\widehat{\Pi}_k\,\big|\, k=1,2, \dots \big\}$, consisting of abstract, mutually disjoint, 
orthogonal projections, $\widehat{\Pi}_k$, with
\begin{equation}\label{PoU}
\widehat{\Pi}_j^{*}=\widehat{\Pi}_j\,, \quad \widehat{\Pi}_j \cdot \widehat{\Pi}_k = \delta_{jk} \, \widehat{\Pi}_k, \,\,\,\forall\, j,k,
\qquad \sum_{k} \widehat{\Pi}_k = \mathbf{1}\,.
\end{equation}
Given an isolated system $S$, let $\mathfrak{E}_S$ denote the family of all projection operators belonging to 
arbitrary potential events. At every time $t$, there is a representation on the Hilbert space $\mathcal{H}$ of $S$,
\begin{equation}\label{Rep}
\mathfrak{E}_{S} \supset \mathfrak{e} \ni \widehat{\Pi} \mapsto \Pi_t\,,
\end{equation}
of all projections $\widehat{\Pi}$ belonging to a potential event $\mathfrak{e}$ by orthogonal projections $\Pi_t$ acting on $\mathcal{H}$, 
for all potential events $\mathfrak{e}\subset \mathfrak{E}_S$. The argument, $t$, of $\Pi_t,$ with $\widehat{\Pi}\in \mathfrak{e}$, is the
time at which the potential event $\mathfrak{e}$ might set in. We assume that the time-dependence of the projection operators 
$\Pi_t$ on $\mathcal{H}$ is determined by the 
\textit{Heisenberg equations}
\begin{equation}\label{Heisenberg-1}
\dot{\Pi}_t= \frac{i}{\hbar} \big[H_S, \Pi_t\big],\qquad \text{ for }\, t\in \mathbb{R}\,,
\end{equation}
see Eq.~\eqref{Heisenberg}. The family of mutually disjoint orthogonal projection operators $\big\{\Pi_t\,\big|\, \widehat{\Pi} \in \mathfrak{e} \big\}$ on $\mathcal{H}$
 is interpreted as the \textit{potential event} (corresponding to $\mathfrak{e}\subset \mathfrak{E}_S$) \textit{possibly setting in at time} $t\in \mathbb{R}$.

We denote by $\mathcal{E}_{\geq t}$ the (weakly closed $^{*}$-) algebra generated by all the operators
\begin{equation}\label{algebra} 
\big\{ \Pi_{t'}\,\big|\, \widehat{\Pi} \in \mathfrak{e}\subset \mathfrak{E}_S, t' \geq t\big\}\,,
\end{equation}
where $t\in \mathbb{R}$ is an arbitrary time. The Heisenberg equation \eqref{Heisenberg-1} implies that all these algebras are isomorphic to one another; 
\begin{equation}\label{Iso}
\mathcal{E}_{\geq t'} = e^{i(t'-t)H_S/\hbar}\, \mathcal{E}_{\geq t}\, e^{i(t-t')H_S/\hbar}, \quad \text{ for arbitrary } \, t, t' \text{ in }\, \mathbb{R}\,.
\end{equation}
From the definition of these algebras it follows immediately that
$$\mathcal{E}_{\geq t'}\subseteqq \mathcal{E}_{\geq t}\,, \,\,\text{ whenever }\,\, t'>t\,.$$
It turns out that, in an isolated \textbf{open} system, a stronger property must hold:
\begin{equation}\label{PDP}
\boxed{\, \mathcal{E}_{\geq t'}\subsetneqq \mathcal{E}_{\geq t}\,, \,\,\text{ whenever }\,\, t'>t\,.}
\end{equation}
This property, called \textbf{``Principle of Diminishing/Declining Potentialities''} (PDP), was first introduced (under the name of
``loss of (access to) information'') in \cite{FS}, with the aim of solving the ``Measurement Problem.'' It can be viewed as a general
characterization of isolated \textbf{open} systems, more precisely of \textbf{dissipation} in isolated open systems. (It may seem that
PDP is somewhat analogous to Sommerfeld's radiation condition; but this is a false impression: PDP is a \textit{purely algebraic} principle 
that does not exclude that photons may hit an isolated open system from the outside. It is expected to hold on a very general stratum of 
inital states.) It turns out that in relativistic Quantum Electrodynamics on even-dimensional Minkowski space-time, with $t$ the time 
of some ``observer,''\footnote{more specifically, \textit{Landau's} ``synchronous time,'' which may be determined by the
\textit{``mimetic field''} introduced in \cite{C-M}} and in limiting theories on Newtonian space-time, $\mathbb{E}^{3}\times \mathbb{R}$, 
obtained by letting the velocity of light, $c$, tend to $\infty$, PDP is a \textbf{theorem}.
It follows from results of D.~Buchholz and J.~E.~Roberts \cite{Buchholz, BR}, who established a fundamental algebraic property 
under the name of ``Huygens Principle'' that implies PDP, and it continues to hold in the limit  $c\rightarrow \infty$, as verified, 
e.g., in \cite{FGP}. It can also be argued to hold on more general space-times in the presence of black holes. But, for systems 
of \textit{massive} matter \textbf{not} coupled to any quantum field with massless modes PDP \textbf{fails}, the algebra 
$\mathcal{E}_{\geq t}$ is \textbf{independent} of time $t$, and the acquisition of information about such systems in 
``measurements'' turns out to be impossible.

\subsection{What are states?}

Next, we introduce a natural notion of \textbf{states}. A state of a system $S$ at time $t$ is supposed to be a mathematical
object enabling one to predict the likelihood or frequency of an arbitrary potential event $\mathfrak{e}\subset \mathfrak{E}_S$ 
to set in at some time $t'\geq t$. This suggests to define a \textit{state} of $S$ at time $t$ to be a normalized positive 
linear functional on the algebra $\mathcal{E}_{\geq t}$ (or, equivalently, a quantum probability measure, in the sense 
of Gleason \cite{Gleason}, on the family of all potential events of $S$ possibly setting in at some time $t'\geq t$). It turns out to 
be necessary to also introduce a notion of \textbf{``ensemble states.''} Let $E_S$ denote an ensemble of very many systems
all of which are isomorphic to a system $S$ and prepared in the \textit{same} initial state at some initial time $t_0=0$. 
An \textit{ensemble state} of $S$ at a time $t>0$ is defined as the average over the ensemble $E_S$ of states at time $t$ of 
systems in $E_S$ that have evolved from the \textit{same} initial state prepared at time $t=0$. Suppose the systems in $E_S$ 
are all prepared at time $t=0$ in a state $\omega_0$ on the algebra $\mathcal{E}_0:= \mathcal{E}_{\geq 0}$. 
In the Heisenberg picture, the \textit{ensemble state,} $\omega_t$, at some time $t>0$, given the initial state $\omega_0$, 
is defined to be the \textit{restriction} of the state $\omega_0$ to the subalgebra $\mathcal{E}_{\geq t}$ of $\mathcal{E}_0$; i.e.,
\begin{equation}\label{ensemble state}
\boxed{\, \omega_t := \omega_0 \big|_{\mathcal{E}_{\geq t}}\,, \quad \text{ for an arbitrary } \,t>0\,.}
\end{equation}

In non-relativistic QM (in the limit where $c\rightarrow \infty$), the algebra $\mathcal{E}_{\geq t}$ is isomorphic to
the algebra of all bounded operators on a Hilbert space, $\mathcal{H}_t$,
\begin{equation}\label{Property P}
\mathcal{E}_{\geq t} \simeq B(\mathcal{H}_t)\,,
\end{equation}
i.e., the algebra $\mathcal{E}_{\geq t}$ is a so-called type I$_{\infty}$-factor, for arbitrary $t\in \mathbb{R}$. 
The isomorphism in Eq.~\eqref{Property P} is called \textbf{Property P}; see \cite{FGP}.\footnote{Eq. \eqref{Property P} 
is \textbf{not} true in relativistic theories with $c< \infty$, in which case the algebras $\mathcal{E}_{\geq t}$ 
are typically von Neumann algebras of type III$_1$.}

\textit{\underline{Remark}:} For $t'>t$, the relative commutant of the algebra $\mathcal{E}_{\geq t'}$ in 
$\mathcal{E}_{\geq t} \,\,(\supsetneqq \mathcal{E}_{\geq t'})$ is then a factor of type I$_{\infty}$, too. Hence there are families 
of Hilbert spaces $\big\{\mathcal{H}_t\,\big|\, t\in \mathbb{R}\big\}$ and $\big\{\mathfrak{h}_{t,t'}\,\big|\, t<t'<\infty\big\}$ 
such that
\begin{align}\label{struct}
\begin{split}
&\mathcal{H}_{t} = \mathfrak{h}_{t,t'} \otimes \mathcal{H}_{t'}\,, \quad \text{for }\,\,t<t'<\infty\,,\,\,\text{with }\\
&\,\,\,\,\mathcal{E}_{\geq t'} =  \mathbf{1}|_{\mathfrak{h}_{t,t'}} \otimes B(\mathcal{H}_{t'}) \subsetneqq B(\mathcal{H}_t) = \mathcal{E}_{\geq t}\,.
\end{split}
\end{align}

Suppose now that $\omega_0$ is a \textbf{pure} state on $\mathcal{E}_0:=\mathcal{E}_{\geq 0}$. Assuming that 
PDP holds, we conclude that, in general, the ensemble state $\omega_t$, for $t>0$, defined in \eqref{ensemble state} is a 
\textbf{mixed} state on the algebra $\mathcal{E}_{\geq t}\subsetneqq \mathcal{E}_0$, a manifestation of \textbf{entanglement}
of the state of the system at time $t$ and the state of fundamentally unobservable modes emitted by the system before time $t$.

This is a crucial observation. It means that the Heisenberg-picture time evolution of \textit{ensemble states} 
introduced in \eqref{ensemble state} is \textbf{dissipative}, converting \textit{pure} states into \textit{mixtures} accompanied
by entropy production. The underlying physical reason for dissipation is that an open system satisfying PDP 
can release modes (``photons'') to the outside world at some time $t$ that, \textit{for fundamental reasons,} escape from the system 
for good and become \textit{unobservable} at any later time $t'\geq t$, causing \textit{decoherence.} In the non-relativistic limit, 
with $c\rightarrow \infty$, such modes escape to spatial infinity infinitely rapidly.

It appears to be a \textit{general fact}\footnote{related to the dissipation-fluctuation theorem, as will be discussed in more detail elsewhere.} -- not only in QM, 
but also in classical physics -- that the dissipative nature of the time-evolution of ensemble states implies that 
the states of \textbf{individual} systems belonging to the ensemble exhibit a \textbf{stochastic} time-evolution
obtained by ``unraveling'' the dissipative (but deterministic) evolution of ensemble states, provided that an appropriate 
``ontology'' of individual systems is assumed.

It may help the readers' understanding of this fact to mention an example in classical physics. 
Imagine that we consider a large ensemble -- a ``gas'' -- of point-like particles suspended in a thermal liquid 
and exhibiting Brownian motion -- caused by collisions of the  particles with lumps in the liquid -- but not interacting 
with one another. An ensemble state of this system at time $t$ is given by the density, $\rho_t$, of particles in 
physical space $\mathbb{E}^{3}$. The time evolution of $\rho_t$ is given by a \textit{diffusion equation,} which is a 
\textit{linear, deterministic} equation known to be \textit{dissipative} and to produce entropy. The evolution of the state
of a single particle in the ensemble, namely its position, $\xi(t)\in \mathbb{E}^{3}$, as a function of time $t$ is, however, 
\textit{stochastic;} it is given by a \textit{Wiener process,} which can be constructed by ``unraveling'' the diffusion equation; 
(see, e.g., \cite{Nelson}). More details about this example and the analogy with QM can be found in \cite{Fr, FGP}.

\subsection{Time-evolution of ensemble states as Lindblad evolution \cite{Lindblad}}

Next, we intend to characterize the time evolution of ensemble states introduced in \eqref{ensemble state} more
explicitly. For this purpose, it is convenient to switch from the Heisenberg picture to the \textit{Schr\"odinger picture.} 
Let $S$ be an isolated system of the kind considered above, and let $\Omega_0$ be the density matrix on the Hilbert space 
$\mathcal{H}_0$ representing the initial state $\omega_0$ of $S$. By Eq.~\eqref{ensemble state}, the density 
matrix $\Omega_t$ representing the ensemble state $\omega_t$ on the Hilbert space $\mathcal{H}_t$, with 
$\mathcal{H}_0=\mathfrak{h}_{0, t} \otimes \mathcal{H}_{t}$, is then determined by restricting the state $\omega_0$ given by the 
density matrix $\Omega_0$ to operators of the form $\mathbf{1}|_{\mathfrak{h}_{0, t}}\otimes Y$, with $Y\in \mathcal{E}_{\geq t}$.
Every operator $Y$ in the algebra $\mathcal{E}_{\geq t}\subsetneqq \mathcal{E}_0$ can be written as 
$\mathbf{1}|_{\mathfrak{h}_{0, t}}\otimes Y= e^{itH_S/\hbar}\, X\, e^{-itH_S/\hbar}$, for some operator $X\in \mathcal{E}_0$. 
We may thus write
\begin{align}\label{state-t}
\begin{split}
\omega_t(Y)\overset{\eqref{ensemble state}}{=}& \omega_0(\mathbf{1}|_{\mathfrak{h}_{0, t}}\otimes Y) 
=\text{Tr}\big[\Omega_0 
\cdot (\mathbf{1}|_{\mathfrak{h}_{0, t}} \otimes Y)\big]\\
=&\text{Tr}\big[\Omega_0 \cdot e^{itH_S/\hbar}\, X\, e^{-itH_S/\hbar}\big] =:\text{Tr}\big[\widetilde{\Omega}_t\cdot X\big]\,,
\end{split}
\end{align}
for all $Y\in \mathcal{E}_{\geq t}$, hence for all operators $X$ in $\mathcal{E}_0$, 
where
\begin{equation}\label{Schrodinger}
 \widetilde{\Omega}_t := e^{-itH_S/\hbar}\,\Omega_0\,\, e^{itH_S/\hbar}
\end{equation}
is again a density matrix on the Hilbert space $\mathcal{H}$ of $S$ whose restriction to the algebra $\mathcal{E}_0$
determines a density matrix, $\Omega_{t}$ on $\mathcal{H}_0$, given by $\Omega_{t}:=\widetilde{\Omega}_{t}\big|_{\mathcal{E}_0}$.
The map $\Omega_0 \mapsto \Omega_t$ determined by Eq.~\eqref{Schrodinger} is \textit{linear, trace-preserving} and 
\textit{positivity-preserving}. It is actually \textit{completely positive}.

The arguments sketched here can be repeated to show that the maps
\begin{equation}\label{Kraus maps} 
\Gamma_{t',t}: \Omega_t \mapsto \Omega_{t'}, \quad t'>t\geq 0\,,
\end{equation}
on the space of density matrices on $\mathcal{H}_0$ are \textit{linear, trace-preserving, positivity-preserving} 
and \textit{completely positive}. One easily verifies that
$$\Gamma_{t', t''} \circ \Gamma_{t'',t} = \Gamma_{t',t}\,, \,\,\,\text{for }\, t'\geq t''\geq t, \quad \Gamma_{t,t} = \text{id }.$$
The advantage of the Schr\"odinger picture is that time-evolution can be viewed as an evolution of density matrices
on the \textbf{fixed} Hilbert space $\mathcal{H}_0$ under linear, trace-preserving, positivity preserving, completely positive
maps. Such maps have been characterized by Kraus \cite{Kraus}.

For simplicity, we henceforth consider a system $S$ of matter interacting with the quantized radiation field in the limiting regime
where $c=\infty$ and with the radiation field prepared in its \textit{vacuum state,} which is time-translation invariant. 
Since, for such a system, ``photons'' created at some time $t$ escape to an event horizon infinitely rapidly, the state of the 
radiation field at any time $t'>t$ is again the vacuum state. It then turns out that the maps $\Gamma_{t', t}$, with $t'>t,$ only depend on the
difference variable $t'-t$, i.e., $\Gamma_{t',t}= \Gamma(t'-t)$, and that the radiation field can be eliminated from our description; 
see \cite{FGP}. Assuming that $\Gamma(t)$ converges strongly to the identity map, as $t\searrow 0$, one concludes that 
$\big\{\Gamma(t)\,\big|\, t\geq 0\big\}$  is generated by a \textit{Lindblad operator,} $\mathcal{L}$, i.e., $\Gamma(t)= e^{t\mathcal{L}}$,
for $t\geq 0$. Thus, the time-evolution of the density matrices $\Omega_t, t\geq 0,$ is governed by a Linblad equation (see \cite{Lindblad})
\begin{equation}\label{Lindblad}
\dot{\Omega}_t = \mathcal{L}\big[\Omega_t\big]\,.
\end{equation}
The general form of a Lindbladian, $\mathcal{L}$, is given by
\begin{equation}\label{Lindbladian}
\mathcal{L}\big[\Omega\big]= -\frac{i}{\hbar} \big[H_{S}^{0}, \Omega\big] + \alpha \sum_{\sigma=1,2,\dots} \Big[T_{\sigma}\,\Omega_t\,T_{\sigma}^{*} -
\frac{1}{2}\big\{\Omega, T_{\sigma}^{*}\,T_{\sigma}\big\}\Big]\,.
\end{equation}
where $H_{S}^{0}$ is a self-adjoint ``Hamiltonian'' on $\mathcal{H}_0$, $\alpha\geq 0$ is (the square of) a coupling constant,  
$\big\{T_{\sigma}\,\big|\, \sigma=1,2,\dots\big\}$ is a family of arbitrary (bounded) operators on $\mathcal{H}_0$, and
$\big\{\cdot,\cdot\big\}$ denotes an anti-commutator.
If $S$ is a system of matter interacting with the radiation field, the operator $H_{S}^{0}$ is the Hamiltonian of $S$ when it is decoupled 
from the radiation field; $\alpha$ is proportional to the square of the elementary electric charge, and the opertors 
$\big\{T_{\sigma}\,\big|\, \sigma=1,2,\dots\big\}$ are determined by \textit{radiative transition amplitudes} of matter; see, e.g., \cite{FGP}.

\textit{\underline{Remark}:} If the velocity of light is kept finite the notion of (ensemble) states must be refined; (ensemble states then turn out to
be indexed by points in space-time, rather than just depending on time). While, as a consequence of ``Huygens' Principle'' \cite{Buchholz}, 
the evolution of ensembles states is shown to be dissipative, Property P (see \eqref{Property P}) fails, and one will encounter memory effects. 
Some general ideas about the $ETH$-Approach to relativistic quantum theory are described in \cite{Fr-2, Fr}; but plenty of details remain
to be understood more precisely.

\section{The stochastic time-evolution of states of \textit{invidual} systems}\label{stoch evol} 

In this section, we set $\hbar=1$. Our aim is to substantiate the claim that whenever the time-evolution of ensemble states is 
\textbf{dissipative}, with pure states evolving into mixtures, then the evolution of states of individual systems is \textbf{stochastic}. 
To begin with, we describe some further ingredients of the $ETH$-Approach to QM that will turn out to be essential.

\textit{\underline{Ontology}:} In the Schr\"odinger picture of non-relativistic QM, as described in Sect.~\ref{dissipation}, 
states occupied by an \textbf{individual} isolated system $S$ at an arbitrary time $t$ are assumed to be given 
by density matrices proportional to finite-dimensional orthogonal projections, $\Pi_t$, in the algebra $\mathcal{E}_0$, i.e.,
\begin{equation}\label{individual system}
\omega_t(X)=\frac{1}{\text{Tr}[\Pi_t]}\text{Tr}_{\mathcal{H}_t}\big[\Pi_t\cdot X\big],\quad \forall\,\, X\in \mathcal{E}_0\,.
\end{equation}
(In the Heisenberg picture, the state corresponding to $\Pi_t$ would be a projection in the algebra $\mathcal{E}_{\geq t}$.)
Generically, the projection $\Pi_t$ has rank 1, i.e., it defines a \textit{pure} state. The assumption in Eq.~\eqref{individual system}
is analogous to the one made in the theory of diffusion of a gas of non-interacting particles alluded to at the end of Subsect.~2.1: 
The state at an arbitrary time $t$ of an \textit{individual} particle in the gas is given by its position $\xi(t)\in \mathbb{E}^{3}$, 
corresponding to a density $\rho_t(x)=\delta_{\xi(t)}(x),\, x\in \mathbb{E}^{3}$, which is a \textit{pure} state; (see \cite{FGP} for
details on the example of diffusion of particles on a simple (hyper-)cubic lattice and the associated theory of simple random walk.)

Let $\Omega$ be a density matrix on $\mathcal{H}_0$ representing a state $\omega$ of $S$ at some fixed positive time. 
The spectral theorem says that
\begin{equation}\label{spect thm}
\Omega= \sum_{\delta=0, 1,2, \dots} p_{\delta} \, \Pi^{\delta}, \qquad p_0 > p_1 > \dots \geq 0\,, 
\sum_{\delta=0,1,2,\dots} \Pi^{\delta}= \mathbf{1}\,,
\end{equation}
with
$$\text{Tr}[\Omega] = \sum_{\delta=0, 1,2,\dots} p_{\delta} \, \text{dim} (\Pi^{\delta}) = 1, \qquad \text{dim}(\Pi^{\delta}):= 
\text{Tr}[\Pi^{\delta}]\,,$$
where the operators $\big\{\Pi^{\delta}\,\big|\, \delta=0, 1,2,\dots\big\}$ are the spectral projections of $\Omega$. They generate
a partition of unity by mutually disjoint orthogonal projections and can therefore be interpreted as representing a 
\textit{potential event.} We note that
\begin{align}\label{state-dec}
\begin{split}
\omega(X)=& \sum_{\delta=0,1,2,\dots} \omega(\Pi^{\delta}\,X\,\Pi^{\delta})\,, \quad \text{ with }\\
\omega(\Pi^{\delta}\,X\,\Pi^{\delta})=& \text{Tr}\big[\Omega\, \Pi^{\delta}\,X\big]= p_{\delta} \text{Tr}\big[\Pi^{\delta} \,X\big]\,, 
\qquad \forall X \in B(\mathcal{H}_0)\,.
\end{split}
\end{align}
If, at some time $t$, the state averaged over a large ensemble $E_S$ of systems, all identical to a system $S$, is given by $\omega$ 
then, according to \eqref{individual system}, the state of an \textit{individual} system in $E_S$ is assumed to be given by the 
density matrix $[\text{dim}(\Pi^{\delta})]^{-1} \Pi^{\delta}$, with a frequency given by $p_{\delta}\, \text{dim}(\Pi^{\delta})$, 
for some $\delta=0, 1,2, \dots,$ as predicted by \textit{Born's Rule.} 
In the $ETH$-Approach to QM, this assumption is called \textbf{``State-Selection Postulate''} \cite{BFS, Fr}. Besides PDP, this postulate 
represents a natural requirement for \textit{isolated open systems.} One then says that the potential event 
$\big\{\Pi^{\delta}\,\big|\, \delta = 0,1,2,\dots\big\}$ \textit{actualizes} at time $t$, and one calls the projection $\Pi^{\delta}$ 
correspondig to the state $\text{dim}(\Pi^{\delta})^{-1}\,\Pi^{\delta}$ occupied by some individual systems in $E_S$ an 
\textit{``actual event''} or \textit{``actuality''}; see \cite{Fr, FGP}.

\subsection{Derivation of the stochastic time-evolution}

We now apply the \textit{State-Selection Postulate} to deduce the stochastic time-evolution of an \textbf{individual} 
system belonging to the ensemble $E_S$ from the time-evolution of ensemble states given in  Eq.~\eqref{Lindblad}; 
(we follow the presentation in \cite{FGP} and use notations similar to those in that paper). Suppose that, at some time $t$, 
all systems in $E_S$ occupy a state $\Pi_t$ (for simplicity assumed to be a pure state, i.e., given by a rank-1 
orthogonal projection). Equation \eqref{Lindblad} then tells us that the ensemble state at a later time $t+dt$ 
(where $dt$ is a small time-increment) is given by a density matrix
\begin{equation}\label{mixed state}
\Omega_{t+dt}= \Pi_t + \mathcal{L}\big[\Pi_t\big] \, dt + \mathcal{O}(dt^{2})\,,
\end{equation}
where $\mathcal{L}$ is the Lindbladian of Eq.~\eqref{Lindblad}. The spectral decomposition of $\Omega_{t+dt}$ 
takes the form
\begin{equation}\label{deco}
\Omega_{t+dt} = p_{nj}[t, t+dt]\, \Pi_{t+dt}^{0} + \sum_{\delta=1, 2, \dots} p_{\delta}[t, t+dt]\, \Pi_{t+dt}^{\delta}\,,
\end{equation}
where $\Pi_{t+dt}^{0}, \Pi_{t+dt}^{1}, \Pi_{t+dt}^{2}, \dots,$ are mutually disjoint orthogonal projections (the spectral projections
of $\Omega_{t+dt}$), with $\sum_{\delta=0,1,2,\dots} \Pi_{t+dt}^{\delta} = \mathbf{1}\big|_{\mathcal{H}_0}$, 
and, assuming that $dt$ is ``small,''
$$ \Pi_{t+dt}^{0} \approx \Pi_t\,.$$
Furthermore,
\begin{align}\label{probabilities}
\begin{split}
&p_{nj}[t, t+dt] \equiv p_{0}[t, t+dt] = 1- \mathcal{O}(dt) >0\,,\\
p_{nj}>&p_1 >  p_2 > \dots \geq 0, \quad \text{with }\,\, p_{\delta} = \mathcal{O}(dt), \,\,\forall\,\, \delta \geq 1\,,\\
\qquad p_{nj}&[t, t+dt] + \sum_{\delta=1,2, \dots} p_{\delta}[t, t+dt] \,\text{dim}(\Pi_{t+dt}^{\delta}) = 1\,,
\end{split}
\end{align}
where ``$nj$'' stands for \textit{``no jump''}. According to the state-selection postulate of the $ETH$-Approach, \textit{one} 
of the density matrices
$$\big\{ \text{dim}(\Pi_{t+dt}^{\delta})^{-1}\Pi_{t+dt}^{\delta}\,\big|\, \delta = 0, 1,2, \dots \big\},$$
\textit{randomly chosen} with a frequency given by $p_{\delta}[t, t+dt]\,\text{dim}(\Pi_{t+dt}^{\delta})$, is the state of an \textit{individual} 
system in $E_S$ at time $t+dt$. The transition from the state $\Pi_t$ to one of the states 
$\big\{\text{dim}(\Pi_{t+dt}^{\delta})^{-1}\, \Pi_{t+dt}^{\delta}\,\big|\, \delta=0,1,2,\dots\big\}$ is the event actualizing 
at time $t$ in an individual system. If $\Pi_t$ evolves to $\Pi_{t+dt}^{\delta}$ in the time interval $[t, t+dt)$, for some $\delta \geq 1$, 
we say that a \textbf{``quantum jump''} is taking place in the interval $[t, t+dt)$.

\subsection{Explicit formulae for the stochastic time-evolution}

In \cite{FGP}, explicit formulae for the projections $\Pi_{t+dt}^{\delta}$ and the coefficients $p_{\delta}[t, t+dt], \delta=0,1,2,\dots,$ 
have been derived from Eq.~\eqref{mixed state}, using a form of analytic perturbation theory dubbed \textit{infinitesimal perturbation theory}; 
see Appendix A of \cite{FGP}. Here we only summarize the results of the analysis in that paper.
\begin{enumerate}
\item[(i)]{Suppose that, during an interval of times $[t_1, t_2)$ there are \textit{no quantum jumps,} in the sense that, for all $t\in[t_1, t_2)$, 
the projection $\Pi_{t}^{0}=: \Pi_t$ describes the state of some individual system in $E_S$, prepared at time $t=t_1$ in a state $\Pi_{t_1}$. 
The probability for this trajectory of states to be observed is given by
\begin{equation}\label{no jumps}
p_{nj}[t_1, t_2]= \text{exp}\Big\{\int_{t_1}^{t_2} \text{Tr}\big[\Pi_t\, \mathcal{L}[\Pi_t]\big] dt\Big\} \leq 1\,,
\end{equation}
where it is used that
\begin{align*}
\text{Tr}\big(\Pi_t \,\mathcal{L}[\Pi_t]\big)&= \text{Tr}\big(\mathcal{L}[\Pi_t]\big) - 
\text{Tr}\big(\Pi_t^{\perp}\, \mathcal{L}[\Pi_t]\big)\\
&= -\alpha \sum_{\sigma} \text{Tr}\big[\Pi_t^{\perp}\,T_{\sigma}\, \Pi_t  \,T_{\sigma}^{*}\,\Pi_t^{\perp}\big]\,\leq 0\,,
\end{align*}
with $\Pi_{t}^{\perp}:= \mathbf{1}- \Pi_t$; the first term on the right side vanishes, while the second term is $\leq 0$, 
by Eq.~\eqref{Lindbladian}.
}
\item[(ii)]{In these formulae, the projection $\Pi_t \equiv \Pi_{t}^{0}$ is a solution of the cubic ordinary differential equation
\begin{equation}\label{time-dep}
\frac{d\Pi_t}{dt}= \Pi_t^{\perp}\, \mathcal{L}[\Pi_t]\, \Pi_t + \Pi_t\, \mathcal{L}[\Pi_t]\, \Pi_t^{\perp}\,,
\qquad  t_1\leq t< t_2\,.
\end{equation}
}
\item[(iii)]{The probabilities $p^{\delta}[t_2,t_2+dt], \delta =1, 2, \dots,$ for observing a ``quantum jump'' during 
the time interval $[t_2, t_2+dt)$ are given by the eigenvalues of the \textit{non-negative} matrix 
$$\Pi_{t_2}^{\perp} \,\mathcal{L}[\Pi_{t_2}]\, \Pi_{t_2}^{\perp}\cdot dt \overset{\eqref{Lindbladian}}{=}
\alpha \sum_{\sigma} \Pi_{t_2}^{\perp}\,T_{\sigma}\, \Pi_{t_2}  \,T_{\sigma}^{*}\,\Pi_{t_2}^{\perp}\, dt\,, \quad \text{i.e.,}$$
\begin{equation}\label{p-jump}
\big\{p^{\delta}[t_2, t_2 +dt]\,\big|\, \delta=1, 2, \dots \big\} = \text{spec}\big(\Pi_{t_2}^{\perp} \,\mathcal{L}[\Pi_{t_2}]\, \Pi_{t_2}^{\perp}\, dt \big)\,,
\end{equation}
up to corrections of order $\mathcal{O}(dt^2)$, with $p^{\delta}[t_2, t_2 +dt] \geq 0, \,\, \forall \,\, \delta=1, 2, \dots$.
}
\end{enumerate}
\textit{\underline{Remark}:} Consider formulae \eqref{no jumps} and \eqref{time-dep} in the special case where matter is
decoupled from the radiation field, i.e., for $\alpha=0$. Then $\text{Tr}\big[\Pi_t \mathcal{L}[\Pi_t]\big] =0\,,$
hence, by \eqref{no jumps}, 
$$p_{nj}[t_1,t_2] \equiv 1\,, \,\,\,\text{ and }\,\, \,\,p^{\delta}[t_2, t_2+dt] = 0,\,\,\forall\,\, \delta=1, 2, \dots,$$ 
for an arbitrary time interval $[t_1, t_2)$; there are \textbf{no} quantum jumps!
Furthermore,
\begin{align*}
\frac{d\Pi_t}{dt} = -i\big\{\Pi_t^{\perp}\,[H_{S}^{0}, \Pi_t] \,\Pi_t + \Pi_t \,[H_{S}^{0}, \Pi_t]\, \Pi_t^{\perp}\big\} = -i[H_{S}^{0}, \Pi_t]\,,
\end{align*}
i.e., for $\alpha = 0$, the time-evolution of the state of the system is described by the Schr\"odinger-von Neumann equation 
with Hamiltonian $H_{S}^{0}$, as expected.

The formulae in Eqs.~\eqref{no jumps} through \eqref{p-jump} uniquely determine a probability measure on the space, $\mathfrak{P}$,
of state trajectories (as functions of time) of individual systems analogous to the \textit{Wiener measure} on the space of
Brownian paths. Explicit expressions can be found in \cite{FGP}.

The theory developed in this section has been applied to the phenomenon of fluorescence in \cite{FGP}. It can also be used
to describe ``measurements'' in quantum mechanics, it predicts in particular the \textit{random time} when a measurement sets in and 
the process of the system's state to jump into a state corresponding to a precise value of the physical quantity that is being 
measured. For models with a discrete time, results on a description of measurements within the $ETH$-Approach 
have been reported in \cite{FP}.

\section{A simple model of a double-slit experiment}
We consider a cavity occupying a cubical domain $\Lambda \subset \mathbb{E}^{3}$ that is composed of two adjacent vacuum chambers
separated by a thin wall, $\Xi$, pierced by two parallel slits and parallel to two opposite faces of $\Lambda$. 
One face of $\Lambda$, denoted by $\Sigma$, parallel to $\Xi$ corresponds to a \textit{light-emitting screen} (a scintillation screen), the opposite face,
denoted by $\Gamma$, of the cavity parallel to $\Xi$ contains the nozzle of a gun emitting quantum particles (electrons or any other kind 
of quantum particles, e.g., Buckminsterfullerenes, interacting with the radiation field), henceforth called ``electrons.'' When an electron hits the light-emitting screen $\Sigma$
a flash of light emanates from a sensitive spot (``pixel'') on $\Sigma$, the location where the electron has hit the screen. 
It is assumed that electrons are emitted by the gun on $\Gamma$ sufficiently rarely that, with very high probability, 
there is either no ``electron'' in the cavity $\Lambda$ or only one electron moving through $\Lambda$ at all times; hence electron-electron 
interactions (many-body effects) can and will be neglected.

The motion of a single electron \textit{decoupled from the quantized electromagnetic field} and \textit{not} interacting
with the pixels on the screen $\Sigma$ is described by a \textbf{Schr\"odinger equation} for the time-dependence
of its state (wave function), with a Hamiltonian given by $-\frac{\hbar^{2}}{2M} \Delta_{\Lambda}$, where $M$ is the mass of the
particle, and $\Delta_{\Lambda}$ is the Laplacian with suitable boundary conditions imposed: On all faces of $\Lambda$, \textbf{except}
$\Sigma$, and on the wall $\Xi$ with the double slit, 0-Dirichlet boundary conditions are imposed on $\Delta_{\Lambda}$. This determines
a densely defined, self-adjoint operator on the Hilbert space $L^{2}(\mathbb{R}^{3}, d^{3}x)$.

Neglecting spin, the total electron Hilbert space, $\mathcal{H}$, is given by
\begin{equation}\label{electron space}
\mathcal{H}= L^2(\mathbb{R}^{3}, d^3x) \oplus \mathfrak{h}_N, \qquad \mathfrak{h}_N\simeq \mathbb{C}^{N}\,,
\end{equation}
where $N$ is the number of states that can be occupied by an electron bound to the light-emitting screen $\Sigma$; these 
states correspond to wave functions localized in the vicinity of the different pixels of the screen $\Sigma$. The idea is that 
an electron that comes close to one of the pixels of $\Sigma$ is attracted to it and tends to jump into a bound-state in 
$\mathfrak{h}_N$ localized near that pixel by emitting a photon that may subsequently be recorded. 

By $P$ we denote the orthogonal projection onto the subspace $L^2(\mathbb{R}^{3}, d^3x)$ of $\mathcal{H}$, and 
$P^{\perp}:= \mathbf{1}-P$ is the projection onto the subspace $\mathfrak{h}_N$. The Hamiltonian of an electron
decoupled from the quantized electromagnetic field and the attractive potentials of the pixels on $\Sigma$ is defined by
\begin{equation}\label{electron Ham}
H_{el}:= -\frac{\hbar^{2}}{2M} P\Delta_{\Lambda} P\,,
\end{equation}

Interactions of an electron  with the light-emitting screen $\Sigma$ are taken into account by the following somewhat idealized 
interaction Hamiltonian
\begin{equation}\label{interaction}
H_I:= g \sum_{\sigma \in \mathfrak{S}} \Big(T_{\sigma}\otimes A^{*}_{\sigma} + T_{\sigma}^{*}\otimes A_{\sigma}\Big)\,,
\end{equation}
acting on the Hilbert space $\mathcal{H}\otimes \mathcal{F}$, where $\mathcal{F}$ is the Fock space of the quantized 
electromagnetic field, $g$ is a coupling constant (proportional to the elementary electric charge), $\mathfrak{S}$ is a family 
of $N$ symbols labeling pixels in a rectangular array contained in $\Sigma$, where photons are emitted from when they are 
hit by an electron, and, for every $\sigma \in \mathfrak{S}$, the operator $T_{\sigma}$ is a radiative transition matrix from the 
subspace $P\mathcal{H} = L^{2}(\mathbb{R}^{3}, d^{3}x)$ of the electron Hilbert space to a state $\big|\sigma\big> \in \mathfrak{h}_N$ 
in the range of the projection $P^{\perp}$, with $\big|\sigma\big>$ the state of an electron bound to the pixel labelled by $\sigma \in \mathfrak{S}$. 
Let $B_{\sigma, r}$ be the ball of radius $r$ in $\mathbb{R}^{3}$ centered at the location of the pixel labelled by $\sigma$, 
and let $P_{\sigma, r}$ be the projection onto the subspace of $P\mathcal{H}$ of electron wavefunctions with support in the 
complement of the ball $B_{\sigma, r}$. We assume that, for all $\sigma\in \mathfrak{S}$, $T_{\sigma}$ is a bounded operator from 
$P\mathcal{H}$ to $P^{\perp} \mathcal{H}$ with the property that
\begin{equation}\label{Decay}
\Vert T_{\sigma} P_{\sigma, r}\Vert \leq K e^{-r/R}\,, \quad \forall\,\, r>0\,,
\end{equation}
where $K$ and $R$ are some positive constants independent of $\sigma$.
The states $\big|\sigma\big>\,, \sigma \in \mathfrak{S},$ are long-lived meta-stable states; for simplicity, we henceforth assume 
that they can be treated as stationary states. Let $Q_{\sigma}$ be the orthogonal projection onto $\big|\sigma\big>$; without loss of 
generality we assume that the projections $\big\{Q_{\sigma}\big| \sigma \in \mathfrak{S}\big\}$ are \textit{mutually disjoint,} 
with $\sum_{\sigma \in \mathfrak{S}} Q_{\sigma} = P^{\perp}$.
Furthermore, $A^{*}_{\sigma}(t)$ is an operator acting on the Fock space $\mathcal{F}$ that creates a ``photon'' at time 
$t$; ($A^{*}_{\sigma}:=A^{*}_{\sigma}(0)$). We assume that, before an electron is released by the electron gun into an initial state, 
$\Psi_0$, contained in the range of the projection $P$ and supported in the region $\Lambda$, the electromagnetic field is 
prepared in its \textbf{vacuum state}, denoted by $\big|\emptyset\big>$. We continue to describe the electromagnetic field 
in a limiting (non-relativistic) regime where the velocity of light, $c$, approaches $\infty$. In this regime, one has that
\begin{align}\label{CCR}
\big< \emptyset \big| A_{\sigma}(t) \cdot A^{*}_{\sigma'}(t') \big|\emptyset \big> =\, & \delta_{\sigma \sigma'}\,\delta(t-t') \,, 
\quad \text{with } \quad A_{\sigma}(t) \big|\emptyset\big> = 0\,,
\end{align}
for all $\sigma, \sigma'$ in $\mathfrak{S}$. When a photon is emitted by an electron hitting a pixel $\sigma$ at some
time $t$ (corresponding to replacing the vacuum state $\big|\emptyset\big>$ of the electromagnetic field by the state
 $A^{*}_{\sigma}(t)\big|\emptyset\big>$) it is recorded by the pixel $\sigma$; subsequently the state of the electromagnetic field 
 \textbf{immediately relaxes back to the vacuum state}.\footnote{In \eqref{CCR}, one may be tempted to replace $\delta_{\sigma \sigma'}$ 
 by matrix elements, $g_{\sigma \sigma'}$, of a more general positive-definite matrix $g$; but this generalization is without interest.}

Under these assumptions, the basic postulates of the \textit{ETH-Approach} to QM enable one to completely \textbf{eliminate} 
the electromagnetic field from the description of the experiment (see \cite{FGP}); and the effective dynamics of electron ensemble 
states is given by a quantum Markov semi-group described, more explicitly, by the Lindblad equation,
\begin{equation}\label{Double-slit dyn}
\dot{\Omega}=-\frac{i}{\hbar}\big[H_{el}, \Omega\big] + \alpha \sum_{\sigma \in \mathfrak{S}}\Big( T_{\sigma}\, \Omega\, T_{\sigma}^{*} - 
\frac{1}{2} \big\{\Omega, T_{\sigma}^{*}T_{\sigma}\big\}\Big)\,,
\end{equation}
where $\Omega$ is a density matrix on $\mathcal{H}$ describing an electronic ensemble state, and $\alpha = g^{2}>0$;
see Appendix A in \cite{FGP}. (In this paper one also finds a detailed discussion of how the ``Principle of Diminishing/Declining 
Potentialities'' (PDP) follows from ``Huygens' Principle'' \cite{Buchholz}, and what this principle says in the limit where 
$c\rightarrow \infty$.)

\subsection{The fate of individual electrons}
Equation \eqref{Double-slit dyn} is linear, deterministic and \textbf{disspative} (for $\alpha > 0$), due to the coupling 
of electrons to the electromagnetic field, which entails the validity of PDP. Quite generally, the dissipative nature of the time 
evolution of ensemble states of a system is crucial for the emergence of actual events, and, in particular, for the success of 
measurements in the system.

We recall that electron states in the subspace $\mathfrak{h}_N \equiv \text{Ran }P^{\perp} \simeq \mathbb{C}^{N}$ are assumed \textbf{not} 
to evolve under the electronic Hamiltonian $H_{el}$; moreover, they do \textbf{not} evolve under the Lindblad evolution of Eq.~\eqref{Double-slit dyn}, 
either (i.e., they are constant in time), as is easy to verify using that $P\cdot P^{\perp}=0$ and $T_{\sigma}\big|_{\text{Ran}P^{\perp}} =0$,
for all $\sigma \in \mathfrak{S}$.
In particular, a pure state given by a projection $Q_{\sigma}= \big|\sigma \big> \big<\sigma \big|, \sigma \in \mathfrak{S},$  does not 
evolve in time; Eq.~\eqref{Double-slit dyn} implies that $\dot{Q}_{\sigma} \equiv 0$, for all $\sigma\in \mathfrak{S}$\,.

According to the \textit{ETH-Approach} to QM, the evolution of an \textbf{individual} electron is \textbf{stochastic}, for $\alpha>0$,
and the law of this evolution is obtained by \textit{``unraveling''} the Lindblad evolution of ensemble states described in \eqref{Double-slit dyn}
in accordance with the State-Selection Postulate of the $ETH$-Approach; (see Subsect.~3.2).

To determine the unraveling concretely, we use that the ranges of the operators  $T_{\sigma}, \sigma \in \mathfrak{S},$ 
are mutually orthogonal to one another and orthogonal to the range of the projection $P$. The State-Selection Postulate 
says that, \textit{at all times}, an individual electron is in a state proportional to a finite-rank projection, in the case 
at hand in a pure state, and that, after each time step $dt$, the pure state is chosen according to a generalized \textit{Born Rule}: 
The probability that, in a time interval $[t, t+dt)$, an electron jumps from a state, $\Pi_t$, in the range of the projection $P$ 
to the state $P_{\sigma}=\big|\sigma\big>\big< \sigma \big|$ (with a photon emitted by the electron from the spot corresponding 
to the pixel labelled by $\sigma$ and subsequently recorded on the screen $\Sigma$) is given by
\begin{equation}\label{prob sigma}
\alpha\, \text{Tr}\big(T_{\sigma}\Pi_t T_{\sigma}^{*}\big)\cdot dt,
\end{equation}
while the probability for the state of the electron to remain in the subspace $P\mathcal{H}$ is given by
\begin{equation}\label{prob P}
\text{Prob}\big\{\text{Ran}(\Pi_{t+dt}) \subset \text{Ran}(P)\big\} = 1- \alpha \sum_{\sigma \in \mathfrak{S}} \text{Tr}\big(T_{\sigma}\Pi_t T_{\sigma}^{*}\big)\cdot dt\,,
\end{equation}
for $dt$ small. This result agrees with the prejudices of the Copenhagen interpretation of QM.

As long as the electron does not jump into a state $\big|\sigma\big>,$ for some $\sigma \in \mathfrak{S}$, the dependence on time $t$ 
of the electron state $\Pi_t =: \Pi^{(nj)}_t$ can be found as follows. Consider operators (proportional to rank-1 projections), 
denoted by $\pi_t$, that solve the equation
\begin{equation}\label{eq state}
\dot{\pi}_t = -\frac{i}{\hbar}\big[H_{el}, \pi_t \big] -\frac{1}{2} \alpha \sum_{\sigma \in \mathfrak{S}} 
\big\{\pi_t, T^{*}_{\sigma}\cdot T_{\sigma}\big\}, \quad \text{ with }\,\,
\pi_{t=0} = \Pi_0 =\big|\Psi_0\big>\big<\Psi_0\big|\,,
\end{equation}
where $\Psi_0$ is the initial wave function of an electron when it is released by the electron gun. Then
$$\Pi^{(nj)}_t := \frac{\pi_t}{\Vert \pi_t \Vert}, \qquad t>0\,.$$
The total probability, $p_{esc}$ , of an electron to \textit{escape} from the cavity without hitting a pixel, as time $t$ tends to $\infty$,
can be calculated by integrating the rate of decay of the probability, $p(t)$, for the state of the electron to remain in the
range of the projection $P$, (thus \textit{perpendicular} to the range of the projection, $P^{\perp}$, onto stationary states 
$|\sigma>\,, \sigma \in \mathfrak{S}$), which is determined by
\begin{equation}\label{time-law}
\frac{\dot{p}(t)}{p(t)} = - \alpha \sum_{\sigma \in \mathfrak{S}} \text{Tr}\big(T_{\sigma}\Pi^{(nj)}_t T_{\sigma}^{*}\big)\,, \quad  t\geq 0\,,
\,\,\text{ with }\,\, p(t=0) = 1\,.
\end{equation}
One then finds that
\begin{equation}\label{escape probability}
p_{esc} =  \text{lim}_{t\rightarrow \infty} \,p(t)\,.
\end{equation}
Electrons escaping from the cavity will not emit photons and hence will not be detected on the light-emitting screen $\Sigma$.
But if $\alpha>0$ then, with a \textit{positive} probability $1-p_{esc}>0$, an electron jumps into \textbf{one} of the states 
$\big|\sigma\big>\,, \sigma \in \mathfrak{S},$ emitting a photon that produces a flash of light on the screen $\Sigma$ 
near the pixel marked with $\sigma$. After many electrons have been released by the electron gun to propagate through 
the cavity $\Lambda$, the light-emitting screen $\Sigma$ shows the interference pattern expected in a double-slit experiment;
indeed, our results appear to agree with what one finds in actual experiments.

Furthermore, the \textbf{time} when an electron arrives at a pixel $\sigma \in \mathfrak{S}$ to jump into a state $\big|\sigma\big>$,
emitting a photon, is a \textbf{random variable} whose law is determined by \eqref{prob sigma} - \eqref{time-law}. 
Equations \eqref{Double-slit dyn} through \eqref{escape probability} determine a \textit{probability measure} on 
the space, $\mathfrak{P}$, of trajectories of states of electrons emitted by the electron gun (somewhat analogous to the Wiener
measure in the theory of Brownian motion). The treatment of the double-slit experiment presented here is an example of a 
\textit{quantum Poisson process,} as studied in more generality in \cite{FGP}. 

One can go on to refine the experiment by, for example, calculating the effect of shining laser light into the cavity resulting in 
the emergence of \textit{electron tracks} (see \cite{BFF}) and in the gradual \textit{disappearance of interference patterns} 
on the light emitting screen.

It deserves to be emphasized that the \textit{success and actual outcome of this experiment is a consequence of the \textbf{dissipative} 
evolution of ensemble states.} Without dissipation, i.e., for $\alpha=0$, electrons would not emit any photons, and the screen $\Sigma$
would remain dark.

A general treatment of measurements in the non-relativistic regime, within the $ETH$-Approach to QM, will be presented
in a forthcoming paper; (see also \cite{FP}).

\textit{Richard Feynman} famously claimed that nobody understands Quantum Mechanics and that the double-slit 
experiment represents the \textit{``only mystery''} at the heart of this wonderful theory. We hope that this note is a 
modest contribution to a better understanding of Quantum Mechanics and to unraveling this ``mystery.''\\

\textbf{Ackowledgements}: We are indebted to Carlo Albert and Henri Simon Zivi for plenty of useful discussions related
to things discussed in this note and to Jakob Yngvason for asking many good questions about our treatment of measurements 
and challenging us to produce a sketch of our ideas and results.\hspace{0.5cm}

JF gratefully and joyfully remembers his innumerable encounters, interactions and joint efforts with his mentors and friends
Israel Michael Sigal and Barry Simon.
\begin{center}
---
\end{center}

\end{document}